\documentclass{aastex}
\usepackage{spr-astr-addons}

\usepackage{multirow}

\shorttitle{SNR N132D}
\shortauthors{Vogt et Dopita}

\begin{document}

\title{The 3D Structure of N132D in the LMC:  \newline A Late-Stage Young Supernova Remnant}


\author{Fr\'ed\'eric Vogt\altaffilmark{1} and Michael A. Dopita\altaffilmark{1,2}}
\email{fvogt@mso.anu.edu.au, Michael.Dopita@anu.edu.au}

\altaffiltext{1}{Mount Stromlo and Siding Spring Observatories, \\ Research School of Astronomy and Astrophysics, \\ Australian National University, Cotter Road, \\ Weston Creek, ACT 2611, Australia. \\ fvogt@mso.anu.edu.au \& Michael.Dopita@anu.edu.au }
%\altaffiltext{2}{Laboratoire d'Astrophysique, \\ Ecole Polytechnique F\'ed\'erale de Lausanne, \\ Observatoire de Sauvergny, 1290 Versoix, Switzerland. \\vogt@gmail.com}
\altaffiltext{2}{Institute for Astronomy, University of Hawaii, \\2680 Woodlawn Drive, Honolulu, HI 96822}

\begin{abstract}
We have used the \emph{Wide Field Spectrograph} (WiFeS) on the 2.3m telescope at Siding Spring Observatory to map the [O~III] 5007\AA\ dynamics of the young oxygen-rich supernova remnant N132D in the Large Magellanic Cloud. From the resultant data cube, we have been able to reconstruct the full 3D structure of the system of [O~III] filaments.  The majority of the ejecta form a ring of $\sim$12~pc in diameter inclined at an angle of $\sim$25 degrees to the line of sight. We conclude that SNR N132D is approaching the end of the reverse shock phase before entering the fully thermalized Sedov phase of evolution. We speculate that the ring of oxygen-rich material comes from ejecta in the equatorial plane of a bipolar explosion, and that the overall shape of the SNR is strongly influenced by the pre-supernova mass loss from the progenitor star. We find tantalizing evidence of a polar jet associated with a very fast oxygen-rich knot, and clear evidence that the central star has interacted with one or more dense clouds in the surrounding ISM.
\end{abstract}


\section{Introduction}\label{Sec:intro}
Amongst the large number of supernova remnants (SNR) known, there exists an elite sub-class - the Oxygen-Rich Young SNR. This class is composed of a handful of objects, located either in our Galaxy or in nearby neighbors. The prototype of this class is Cas A \citep{Kirshner77,Chevalier78}, and the other members are Puppis A  \citep{Winkler85} and G292+1.8  \citep{Goss79,Murdin79} in our Galaxy, N132D \citep{Danziger76} and 0540-69.3 \citep{Mathewson80} in the LMC and 1E 0102.2-7219 \citep{Dopita81} in the SMC. Another is located in NGC~ 4449  \citep{Kirshner80}. In addition, five potential candidates have been recently identified in M83 \citep{Dopita10}.

In these SNR, optical filaments are detected with high radial velocity - of the order of several thousands km~s$^{-1}$. They display strong emission of $\alpha$-process elements, such as oxygen, neon, argon, sulfur and calcium. However, they do not contain any traces of hydrogen or helium. The general understanding is that these filaments represent helium burnt material ejected from the inner zones of the progenitor star. After shock breakout at the time of the supernova event, those ejecta cool, initially adiabatically and later radiatively in a thermally-unstable manner. This process produces dense knots of ejecta at low temperature which move ballistically away from the SNR center for as long as the swept up material remains small \citep{reynolds2008}. Eventually, these knots pass through, and are shocked by, the reverse shock generated when the primary blast-wave starts to sweep up appreciable amounts of the surrounding interstellar medium (ISM).  Such young SNR are of great interest, as the oxygen-rich clumps directly sample the nuclear-burnt material from deep within the progenitor star (\emph{e.g.} \citet{Lasker91,Morse03}). 

A number of theoretical models have been developed  to simulate the line emission from those shocked oxygen-rich clumps, led by \citet{Itoh81}, and developed further to include heavy elements other than oxygen by \citet{Dopita84b} and \citet{Sutherland95} also added the effects of time-dependent cloud shock emission and photoionization, as well as the bow shock emission produced once the dense oxygen-rich clouds enter the reverse shock. \citet{Borkowski90} showed that electron conduction may well be an important factor in determining the emission line spectrum, since the high oxygen abundance leads to very short cooling timescales and therefore steep temperature gradients in the cloud shocks of these fast-moving knots. The importance of this effect was emphasized by \citet{Docenko10} in their attempts to understand the IR line spectra of the fast-moving knots of Cas A. They found that once the charge exchange processes in the post-shock photoionized region had been correctly taken into account, they could reproduce most of the relevant spectral line ratios within the framework of a single-temperature model for this region.

The dynamics of the fast-moving oxygen rich knots can be regarded as remaining ballistic from the time of their ejection. The reason for this is that the individual oxygen-rich cloudlets are very dense, hence the cloud shock velocities are small $\sim 100$ km~s$^{-1}$ - much smaller than the 1000-2000 km~s$^{-1}$ space velocities of the oxygen-rich knots. In addition, much of the emission comes not from the cloud shock, but from a photoionized precursor (R-Type ionization front) driven by the EUV radiation of the cloud shock \citep{Sutherland95}. The photoionized precursor material  is moving with the original ballistic velocity of the cloudlet. 

The ionization front will pass through the cloudlet in a timescale comparable with the dynamical timescale, the time needed for the cloud to fully enter the reverse shock. The dynamical timescale is itself a few times shorter than the cloud crushing and shredding timescale. For this reason,the radiative lifetime of any individual knot is short. For cloud densities $\sim 100\,\mbox{cm}^{-3}$, it is likely comparable with the 25 year $e-$folding lifetimes observed for the Cas~A knots by \citet{Kamper76}. However, due to the relatively low spatial resolution of the data presented in this paper, such small cloudlets cannot be observed. Instead we see larger complexes of cloudlets, which can survive over much longer periods.

The structure of the clumps, and their distribution around the SNR, is linked both to the physics of the SN explosion itself, and to the density structure of the surrounding ISM. Instabilities, asymmetries or the presence of jets may all play their part. Asymmetries in the spatial distribution of the ejecta may also highlight asymmetries in the reverse shock imposed by a ISM density gradient.  For example, in the case of Cas A, the kinematic study by  \citet{Reed95} found that the ejecta appear to be expanding from a centre displaced from the centre of the bubble, and that there are compositional differences in that the fast-expanding ejecta concentrated on the far side of the SNR show more S, while the near side is relatively richer in O. This may be related to the clear evidence of the formation of a high-velocity jet seen at optical wavelengths \citep{Fesen01,Fesen06a,Fesen06b}. Here, very fast N-rich, S-rich and O-rich knots are seen expanding mainly in the plane of the sky.  In the case of SNR 1E 0102.2-7219 \cite{Vogt10a} were able to deduce that the asymmetries seen in the ejecta were mostly due to an asymmetric bipolar explosion, rather than through any density gradients in the surrounding ISM. 

The SNR studied in this Paper, SNR N132D \citep{Davis76}, is located very close to the bar of the LMC, and almost at the rotation centre of this galaxy. It is known to contain several oxygen-rich filaments with high radial velocities  \citep{Danziger76,Lasker78}. It is amongst the brightest X-ray sources in the LMC, and has been studied numerous times at those wavelengths \citep[e.g.][] {Clark82,Hwang93,Favata97,Hughes98,Borkowski07,Xiao08}. Using observations from the \emph{BeppoSAX} X-ray satellite \citep{Boella97}, \cite{Favata97} suggested that N132D has, or will soon, enter the Sedov phase. The high-resolution X-ray images reveal the position of the forward shock with a shape rather reminiscent of that of a horseshoe \citep[e.g.][]{Xiao08}. \cite{Chen03} modeled this as a blast wave hitting a somewhat irregular cavity wall. This outer shell was also detected in hot dust emission at 24 $\mu m$ by \cite{Tappe06} using the \emph{Spitzer Space Telescope} \citep{Werner04}.The non-thermal radio structure of the outer shock was mapped by \cite{Dickel95} at a wavelength of 6~cm.

The first attempt to dynamically map the three dimensional shape of the oxygen-rich filaments was by \cite{Lasker80}. He described the structure as that of a thin inclined ring. Later on \cite{Morse95} argued that they form, rather, a thin shell. There are no proper motion measurements for the ejecta, unlike other young SNR such as 1E 0102.2-7219 for example \citep{Finkelstein06}. As a result the various age estimates have been based on radial velocities (interpreted as expansion) and on the angular extent of the filament system. Age estimates range from $\sim$1300 years for \cite{Danziger76, Lasker80} to $\sim$2350 years for \cite{Sutherland95a} up to $\sim$3500 years for \cite{Morse95}. The poor agreement between these different studies is a direct consequence of different assumptions about the shape of the O-rich ejecta. Based on \emph{Hubble Space Telecope} FOS UV/optical spectra of the O-rich knots, \cite{Blair00} remarked upon the similarity of their composition with the predictions of the core-collapse supernova nucleosynthesis models of \cite{Nomoto97} for progenitor masses of the order of 25-35~M$_{\odot}$. N132D might therefore be the result of a Type Ib SN. 

Several other clumps located within the outer X-ray shell have been observed to contain hydrogen and other light elements. Since these display an abundance similar to the LMC, it is certain that these represent clouds of pre-existing ISM now undergoing a radiative cloud shock phase. In particular, \cite{Blair00} failed to find any nitrogen enhancement that would suggest a possible origin from mass loss in a stellar wind from a massive progenitor star. Those ISM clumps display radial velocities of the order of a few hundred km~s$^{-1}$, consistent with the expected cloud shock velocities. 

In this paper, we present a new optical study of N132D. Our objective is to remove the uncertainties linked to the shape of the O-rich ejecta by creating a real 3D map of the ejecta \emph{c.f.}  \cite{Vogt10a}. We used the \emph{Wide Field Spectrograph} (WiFeS) on the 2.3m telescope operated by the Australian National University at the Siding Spring Observatory. The instrument design has been described by \citet{dopita2007} and its on-telescope performance has been detailed in \cite{dopita2010}. This instrument provides a (fully filled) field of view of $38\times25$~arc~sec., a spatial resolution element of $1.0\times 0.5$~arc~sec, spectral resolutions of either 3000 or 7000, and a wavelength coverage of 3200-9600 {\AA}. Our analysis is based on the dynamics of the  [O~III] 5007\AA\ emitting filaments at a resolution of $R=3000$.

The data is presented in Section~\ref{Sec:obs}, and the reduction processes are described in Section~\ref{Sec:reduc}. Section~\ref{Sec:results} presents the results, and our 3D map of the oxygen rich ejecta in N132D in is given in Section~\ref{Sec:3d}. We discuss the interpretation of the observations in Section~\ref{Sec:RK}, and our conclusions are presented in Section~\ref{Sec:concl}.

\section{The Observations}\label{Sec:obs}
We obtained spectroscopic maps of the supernova remnant N132D at a resolution $R = 3000$ in the blue and $R = 7000$ in the red.  Data were acquired over four consecutive nights on December 12th to 15th, 2009, using WiFeS (Wide Field Spectrograph, see \cite{dopita2007,dopita2010}) on the ANU 2.3m telescope at the Siding Spring Observatory. The data consist of a mosaic of six fields covering the central and western regions of N132D. Each field spans 25$\times$38 arc sec. with 0.5 arc sec. sampling in the spatial direction. The data were subsequently interpolated to provide $0.5 \times 0.5$ arc sec. pixels. The total field of view covered spans $\sim$81$\times$73 arc sec. in RA and in Dec, respectively. 

\begin{figure*}[htb!]
\centerline{\includegraphics[scale=0.6]{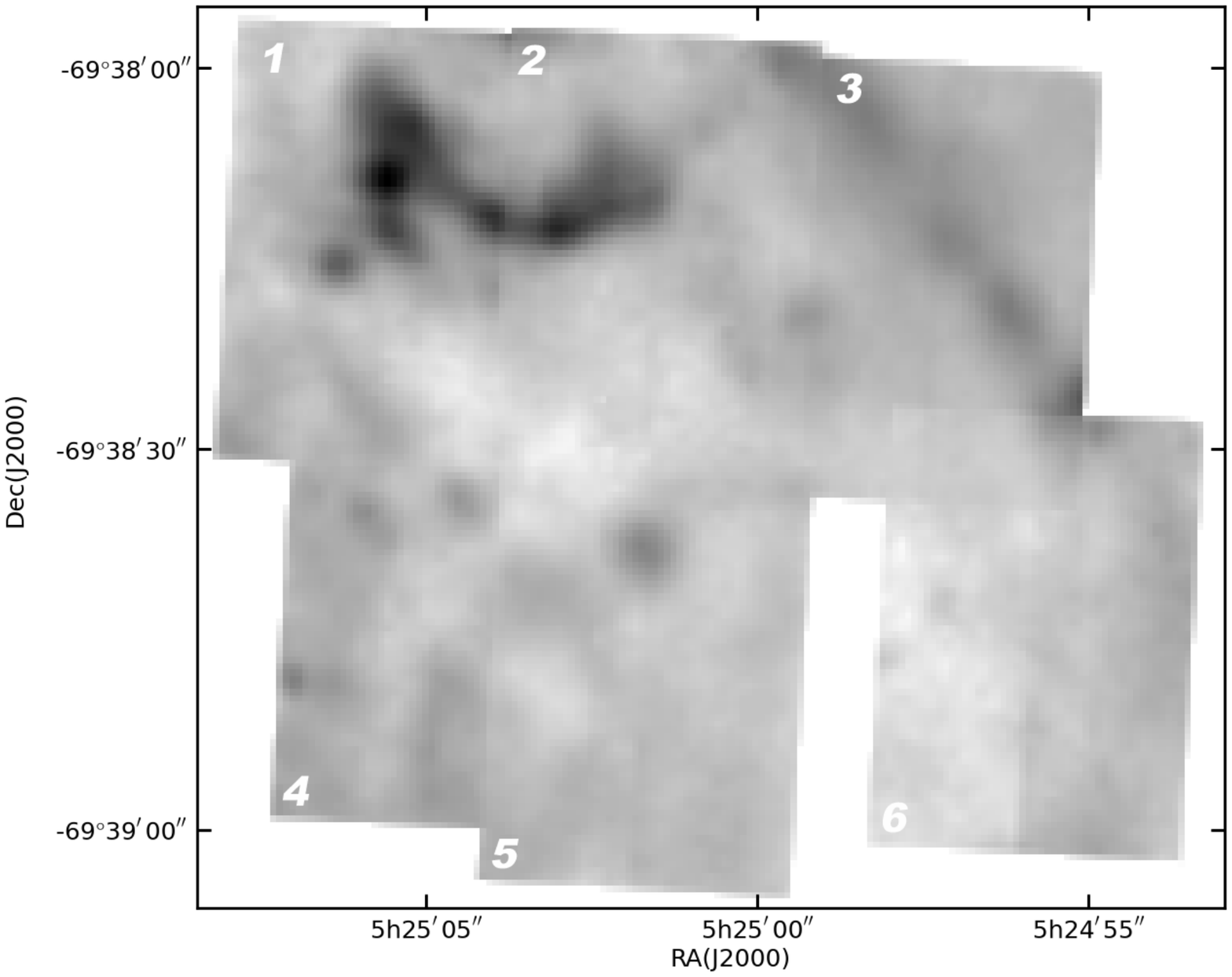}}
\caption{The position, orientation and labeling of the six fields observed in N132D. The images are of the gas in the velocity range -100 to +100 km~s$^{-1}$. The \emph{Lasker's Bowl} \citep[][]{Morse96} is seen to the top in this image of the [O~III] $\lambda$5007 forbidden line.}\label{fig:fields}
\end{figure*}

An observational difficulty of observing the LMC over long periods of the night is that the instrument does not have an atmospheric dispersion compensator. As we have to observe at fixed position angle, the parallactic angle continuously rotates. In addition, the pointing model for the telescope was imperfect at the time of the observation. Because of errors and drift in the pointing and the changes in atmospheric dispersion, the six fields did not properly overlap. In particular, a noticeable gap is present to the bottom right of the mosaic. Following the reduction of the data, described hereafter, we established the pointing of each field \emph{a posteriori} by referencing the star fields to the DSS images of the same area. The accurate pointing position deduced for each field and their respective labeling is shown in Fig.~\ref{fig:fields}. 

The fields numbered 2,5 and 3 were acquired on the nights 1,2 and 3 respectively, with a poor mean seeing of $\sim$3.5, $\sim$3 and $\sim$ 3.5 arc sec. The remaining fields 1,4 and 6 have been obtained in this order on the fourth night, with the seeing ranging from $\sim$1.5 arc~sec. at the beginning of the night to $\sim$ 2.5 arc sec. at the end. Note that the relative spatial shift of field 6 is the worst, mostly resulting from the (large) effect of atmospheric dispersion. The spectral coverage ranges from $\lambda$3200 {\AA} to $\lambda$5900 {\AA} for the blue image, and $\lambda$5290 {\AA} to $\lambda$7060 {\AA} for the red image. Both the red and blue images are taken simultaneously, using the RT560 dichroic which has a cut at about 5700  {\AA} \cite{dopita2010}. 

All images were acquired with $2\times1$ pixel binning using the WiFeS ``Nod-and-Shuffle" data acquisiton mode \citep[see][]{dopita2010}. Each image consists of 6 cycles of 150 sec. on the object and 6 interleaved cycles of 75 sec. on the sky, giving a total on-object exposure of 1350 sec and a total image exposure time (including the overheads for guide-star acquisition and readout) of about 2740 sec. Three such images of each of the six fields were taken to allow for adequate cosmic ray subtraction. 

\section{Data reduction}\label{Sec:reduc}

\subsection{Individual fields}
The reduction procedure followed is identical to that described in \cite{Vogt10a} for their observations of SNR 1E 0102.2-7219. The six fields were each reduced separately using the WiFeS reduction pipeline based upon IRAF \footnote{ IRAF is distributed by the National Optical Astronomy Observatory, which is operated by AURA, Inc., under cooperative agreement with the National Science Foundation.} scripts. This procedure has been described in \citet[][]{dopita2010} and will not be repeated here. 

The same calibration files were used for all fields. The frames used for flat fielding and tracing the slit positions consist of a dome wire and a sky wire (obtained using the coronographic science aperture with a 1.0~arc~sec. wire stretched across the middle), a dome flat and sky flats (obtained using the science aperture. All sky flats were taken at dusk of the first night. Bias and  Ne Ar arc exposures for wavelength calibration were taken  directly before or directly after the group of three exposures on each field . 

Each of the three images of each of the six fields are bias subtracted, flat fielded and sky subtracted, atmospheric dispersion corrected, un-binned, and finally combined and reduced to a wavelength calibrated 3D data cube from which the spectra of individual regions can be extracted, or from which monochromatic images of the whole field can be formed. We have not calibrated the data to absolute flux units using observations of photometric standard stars, as this is not required for the purposes of our dynamical study. 

The issue of the bias subtraction deserves a special mention. The WiFeS bias frames are subject to small temporal variations, but have little or no fixed pattern noise. Since each chip is read out in all four on-chip amplifiers, each quadrant of the bias frame may vary in intensity, and also with respect to the others, as well as suffering small changes in the slope of the mean intensity across the chip. Such variations prevent us from constructing a Master Bias using the standard median techniques. Instead, our Master Bias consists of four smooth planes (one for each quadrant), each one representing the best fit to an individual bias frame. This technique, developed by Jeffrey Rich of the University of Hawaii (private communication) shows very little variation in the bias-subtracted data depending on the time lapse between the Bias acquisition and the Science frame acquisition, and it produces reduced images of better quality when compared to other more standard techniques of bias subtraction. 

\subsection{Merging the fields into a Mosaic}
The six fields were merged into a Master Cube to form the final mosaic. First order alignment of the fields simply used the information contained within the header. We then confirmed the orientation of the separate fields using the position of four stars found in the 2MASS database \citep[see][]{mass2006} that we could clearly identify in our data set. 

Due to the zero-point rotation between the entrance aperture of the spectrograph, and the position angle indicated by the instrument rotator, the six fields are rotated in Position Angle by $\sim$3.2 degrees with respect to the North direction. The slightly fuzzy edges of the fields, that can be seen in Fig.~\ref{fig:fields}, result from the linear interpolation we have used in the rotation of the image cube. 

Once the initial orientation had been determined, the fine alignment of the fields was made manually by aligning the six fields using the DSS image of the area. The final alignment has a notational accuracy of 1 pixel, or 0.5 arc sec. It may in reality be slightly larger than that due to the rather poor seeing we experienced.
   
The five first and five last lines of WiFeS images display high signal variations due to vignetting, atmospheric dispersion, and errors in determination of the slit positions on the original images. Those regions were trimmed for each field before building the mosaic. The final mosaic thus spans $\sim 73$ arc sec. along the Declination direction. In the overlap regions, the pixel value is taken as the median of the overlapping pixel values. An absolute flux calibration was not required. However, we scaled the fields 3 and 6 by a factor of 0.7 and 0.9, respectively, in order to correct for varying atmospheric transmission effects. These factors were determined by matching the signal intensity in the overlapping regions of the data cube. Poor quality of some of the Ne Ar reference arcs introduced a calibration error of one spectral pixel ($\sim$1 {\AA}) for the fields  2,3,5 and 6. This has been corrected manually by reference to the [O~III] $\lambda$ 5007 {\AA}  line emitted in the surrounding ISM, clearly visible in Figure \ref{fig:spectrum}.

\section{Results}\label{Sec:results}

\subsection{De-blending procedure}
Our analysis is based on the dynamics of the [O~III] $\lambda$ 5007 {\AA} forbidden line, which is part of the $\lambda\lambda$ 4959,5007 {\AA} doublet. In the similar study of SNR 1E 0102.2-7219 in the SMC \citep[][]{Vogt10a}, the large velocity dispersion of the O-rich ejecta caused the two emission lines to overlap and blend in some regions of the SNR. In the case of N132D, there is no such overlap, as the velocity dispersion is smaller, and the knots display less spatial overlap. However, despite the lack of direct overlap, we will apply the same \emph{de-blending} procedure for SNR N132D  as for 1E 0102.2-7219. Doing this, we avoid any confusion between $\lambda$~5007 {\AA} blue-shifted and $\lambda$~4959 {\AA} red-shifted knots.

To de-blend, we suppress one of the line group in the [O~III] doublet using the known (theoretical) ratio of 4959 to 5007 {\AA} lines; 1:2.86 \citep[e.g.][]{Dopita2003}. We simply scale down the spectrum by 1/2.86, blue-shift it by 48 {\AA} and subtract it to the original spectrum. This procedure will remove all the $\lambda$4959 contribution from the region of the spectrum $\lambda \lambda$4959-5007. The de-blended spectrum should not be trusted at wavelengths smaller than 4959 {\AA}, as the method will create artifacts in these regions. Fortunately, there is no case in which the radial velocity of the ejecta exceeds -2800 km~s$^{-1}$, sufficient to shift the 5007 {\AA} emission shortward of 4959 {\AA}. One knot, referred to as \emph{B2} \citep[][]{Morse95} hereafter, is close to this threshold, but fortunately does not appreciably exceed it. 

\begin{figure*}[htb!]
\centerline{\includegraphics[scale=0.7]{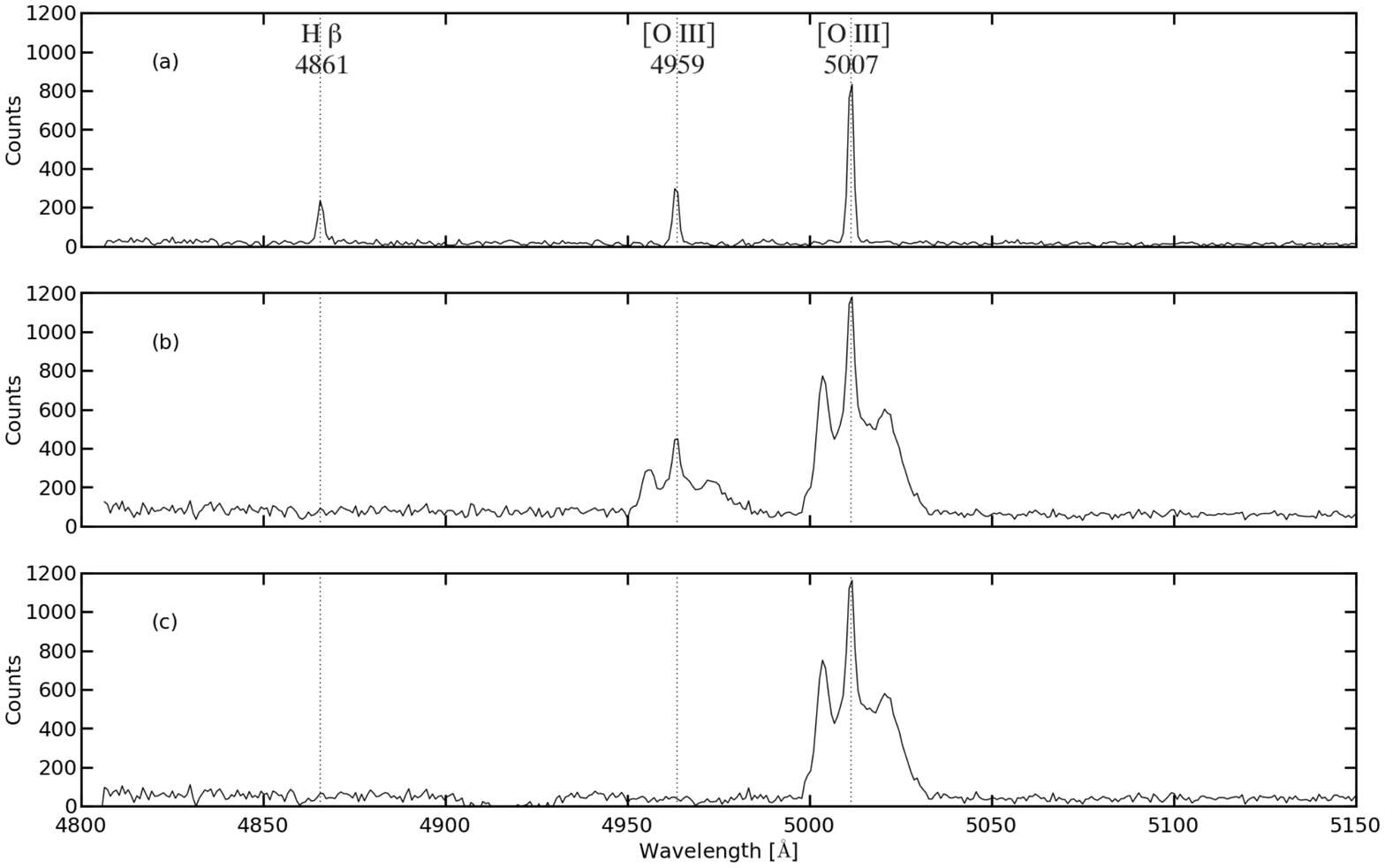}}
\caption{(a) A typical spectrum of the medium surrounding the SNR, showing the narrow H$\beta$ and [O III] $\lambda\lambda$ 4959, 5007 {\AA} lines. These sharp lines are shifted with respect to the rest frame wavelength, due to the radial velocity of the LMC. (b) A typical spectrum of a region containing O-rich knots. The blue- and red-shifted lines are clearly visible. In panel (c) we present the same spectrum with the contribution of the [O III] $\lambda$ 4959 {\AA} line removed using the procedure described in the text. This provides the line profile used in the kinematical analysis.}\label{fig:spectrum}
\end{figure*}

In Fig.~\ref{fig:spectrum}, we display a characteristic spectrum of the surrounding medium of SNR N132D (panel a), a spectrum of the SNR O-rich ejecta prior to de-blending (Figure \ref{fig:spectrum} panel b), and a de-blended spectrum in Figure \ref{fig:spectrum} panel (c). 
The wavelengths of the H$\beta$, and  [O~III] $\lambda\lambda$ 4959, 5007 {\AA} lines arising from the surrounding ISM are marked in dotted lines on each of the panels of this figure. The widths of these lines are largely determined by the instrumental resolution, since the local turbulence in the interstellar gas is low. Their measured wavelengths are respectively 4865.7, 4963.5 and 5011.4 {\AA}. Comparing with their respective rest frame wavelength of 4861.3, 4958.9 and 5006.7 {\AA} \citep[e.g.][and references therein]{Groves02, Dopita2003}, we deduce the heliocentric radial velocity for the interstellar gas surrounding the SNR to be 275$\pm$4 km s$^{-1}$.  Our measured LMC radial velocity is consistent with the previously observed value of $\sim$260 km s$^{-1}$ \citep[see, e.g.][]{Staveley-Smith97,Marel02}. 

Clearly, the de-blended spectrum Figure \ref{fig:spectrum} panel (c) is free of any 4959 {\AA} [O III] line contribution. The main issue here though is that the H$\beta$ line is also affected by this procedure. In our dynamical analysis presented in Section~\ref{sec:orich}, we have used the raw data cube to examine the velocity structure in the H$\beta$ line, and the de-blended cube for the analysis of the [O~III] line.

\subsection{Velocity Map}\label{VMap}
\begin{figure*}[htb!]
\centerline{\includegraphics[scale=0.65]{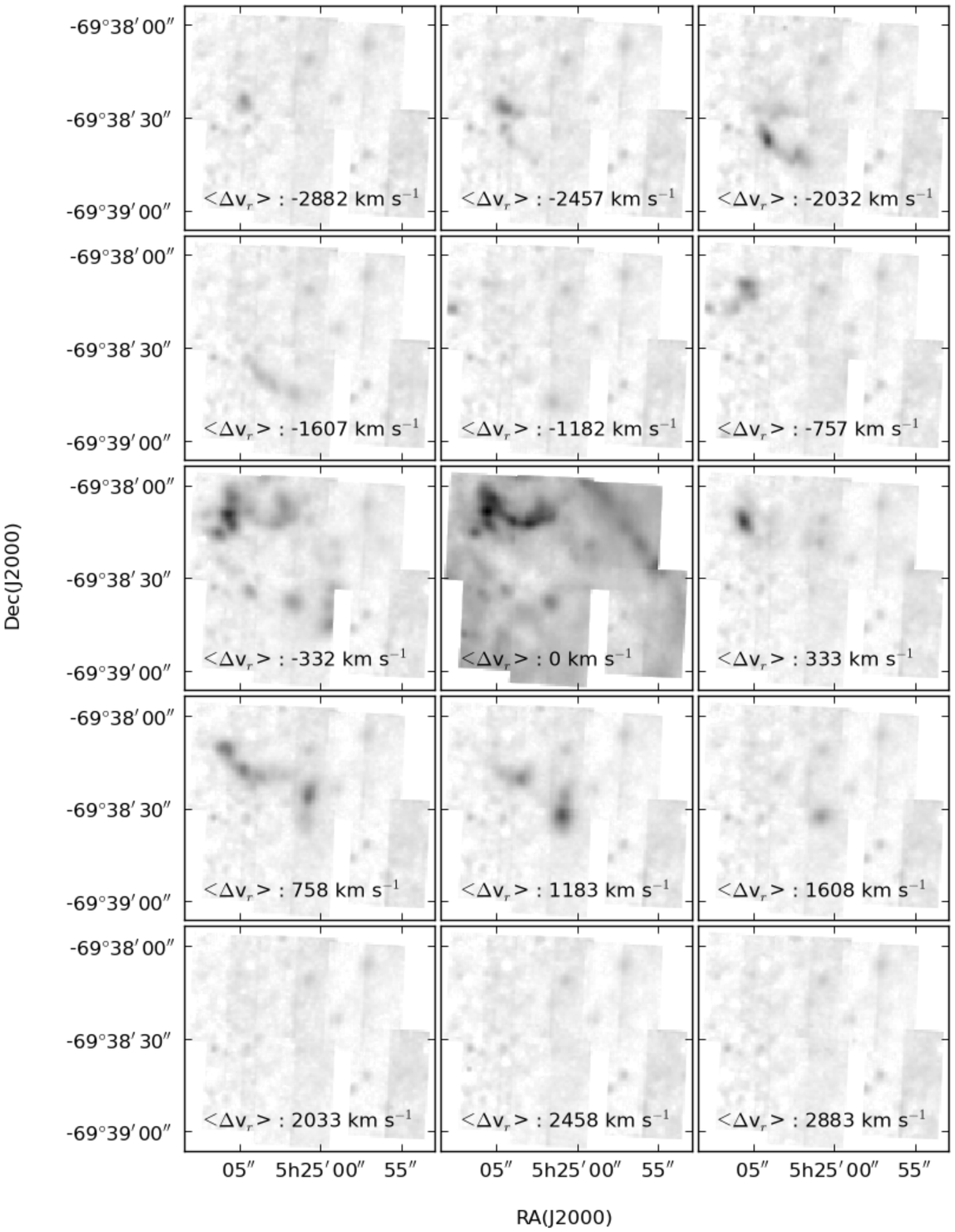}}
\caption{Velocity map of SNR N132D based on the [O~III] forbidden line. Each panel covers a velocity range of $\Delta v_r$=425 km~s$^{-1}$, with exception to the zero velocity panel, which covers a range of $\Delta v_r$=240 km~s$^{-1}$. The mean velocity of each panel is indicated for reference. The image density is proportional to the logarithm of the normalized number of counts to bring out fainter nebulous features. This also brings out artefacts at the $\sim 1-2 $e$^-$pixel$^{-1}$ caused by errors in determining the bias in these quad-readout frames.}\label{fig:vmap}
\end{figure*}

In Fig.~\ref{fig:vmap} we present a velocity map of SNR N132D using the [O~III] $\lambda$ 5007 forbidden line. Each panel covers a velocity range of $\Delta v_r$=425 km~s$^{-1}$, with exception to the zero velocity panel, which covers the smaller range of $\Delta v_r$=240 km~s$^{-1}$. Several individual O-rich knots can be identified in the various panels. In the zero-velocity panel, the so-called \emph{Lasker's Bowl} \citep[][]{Morse96} feature associated with the interaction of the SNR with a large cloud in the surrounding ISM is clearly seen in the NE sector of the remnant (the extended and very bright curved structure). The western rim of the remnant associated with the blast wave can be identified to the NW. 
  
As for the other panels, blue-shifted material is detected with velocities up to $\sim$-2800 km~s$^{-1}$, while red-shifted velocities range up to $\sim$+1800 km~s$^{-1}$. The different clumps tend to trace elongated filaments with a small curvature along them. In that respect, the Lasker's Bowl feature at $<\Delta v_r>$= 0 to 332 km~s$^{-1}$ is an exception - being of comparatively larger size and unconnected with the system of high velocity filaments. 

The O-rich clumps do not cover the whole velocity space. On the blue side especially, the panel with $<\Delta v_r>$= -757 km~s$^{-1}$ shows a gap in the the distribution of strong emitting knots, while they are apparent once again in the panels  with $<\Delta v_r>$= -1182, -1607, -2032 and  -2457 km~s$^{-1}$. There exists a general trend for the O-rich filaments to spread along the  NE-SW direction. In the rest of this paper, we will adopt the labeling convention of \cite{Morse95} to refer to the various O-rich clumps. These knots are labelled with the names given by \cite{Morse95} in the left-hand panel of Fig.~\ref{fig:above}.

Although the oxygen-rich ejecta show a systemic blueshift of $\sim -400$ km~s$^{-1}$, this is not thought to be significant. The instantaneous and accidental spatial distribution of the clumpy ejecta that happen to be passing through the reverse shock will determine whether there is a systematic line shift. Since the fast-moving clumps do not cover the whole velocity space, it is clear that their spatial distribution is very patchy and incomplete.

\subsection{Identifying the O-rich knots}\label{sec:orich}
In the velocity map of Fig.~\ref{fig:vmap}, \emph{all} [O~III] $\lambda$5007 emission is plotted. However, several knots within N132D have already been identified as also containing hydrogen \citep[e.g.][]{Morse96}. Unlike some other young SNR, such as 1E 0102.2-7219 in the SMC \citep{Vogt10a}, SNR N132D appears to be experiencing strong interactions with a highly clumpy surrounding interstellar medium. Whether these clumps are fragments of the ISM which surrounded the supernova progenitor at its birth, or whether they arose from stellar mass-loss and subsequent interactions with a fast pre-supernova Wolf-Rayet wind, or a combination of the two, is yet to be determined. In the X-ray, the clumpy nature of these interactions can be seen in the broken structure of the outer shell \citep[e.g.][]{Borkowski07,Xiao08}. Here, we have used  the presence of H$\beta$ emission to identify and locate knots excited by ISM cloud shocks rather than being the result of high-velocity stellar ejecta. This line is conveniently located close from the [O~III] $\lambda$ 5007 line and can only be emitted in the shocked ISM, not the oxygen-rich ejecta. 

\begin{figure*}[htb!]
\centerline{\includegraphics[scale=0.6]{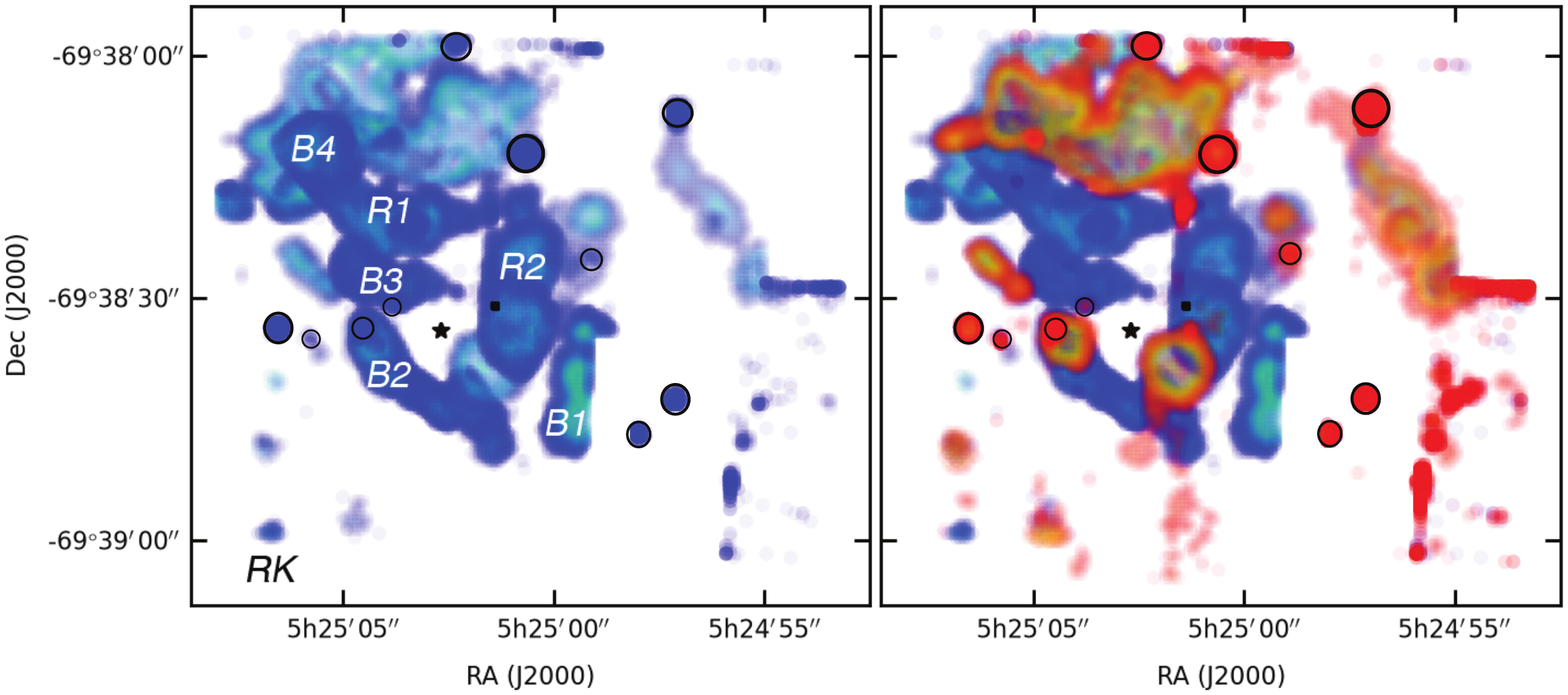}}
\caption{Left : [O~III] forbidden line map of N132D. The zero velocity emission, with -100 km~s$^{-1}$ $\leq$ v$_r$ $\leq$ 100 km~s$^{-1}$ has been removed to avoid filling the figure. The blue-green color ramp is linear in the number of counts per pixel, ranging from 18 to 60. The black star and black square mark the center of the O-rich knots assuming a symmetric distribution, and the remnant center determined from fitting an ellipse to the diffuse outer rim \citep[see][for details]{Morse95}. Right : as the left panel, but including the H$\beta$ emission which is plotted in with a linear color ramp, with pixels counts ranging from 18 (red) to 60 (yellow). See in the text for details. The major knots from \citet{Morse95} have been identified, as well as the run-away knot (RK) discussed in the text, and field stars have been identified as black rings.}\label{fig:above}
\end{figure*}

In the left hand panel of Fig.~\ref{fig:above}, we plot the oxygen emission. In this plot, the intensity scale is scaled linearly from 18 counts per pixel (blue) - up to 60 counts per pixel (green). Low velocity gas, with radial velocity -100 km~s$^{-1}$ $\leq$ v$_r$ $\leq$ 100 km~s$^{-1}$ has been removed in order avoid filling the plane and obscuring the O-rich knots. The  \cite{Morse95} names of each of the O-rich knots are also shown. The right plot also includes the H$\beta$ emission using a red-yellow color ramp. For this too, the counts per pixel range from 18 to 60. Each point is given a transparency of 0.1, so as to render visible superimposed knots. We have also identified the field stars from the continuum image as black rings, to avoid confusion with small emission knots.  We also plot for information the two suggested centers of the SNR from \cite{Morse95} : the black star, located at ($\alpha=5^h25^m02.^s7$; $\delta=-69^{\circ}38^\prime34^{\prime\prime}$), marks the center of the O-rich knots assuming a symmetric distribution. The black square, located at ($\alpha=5^h25^m01.^s4$; $\delta=-69^{\circ}38^\prime31^{\prime\prime}$), marks the remnant center determined from fitting an ellipse to the diffuse outer rim. 

It is now very simple to sort out the true oxygen-rich knots from the shocked ISM clumps. Very prominent as a member of these ISM clouds is the extended \emph{Lasker's Bowl} structure and the outer western edge of the remnant. This structure can only be interpreted as a shocked dense pre-existing interstellar cloud, which has been compressed and shaped by the pre-supernova UV photo-ablation and stellar mass-loss. This would explain its smooth and curved ``inner'' boundary - a classical bow-shock morphology - facing in the direction of the now exploded star.
A number of smaller spherical clumps can also be identified in the southern part of the field. As for the oxygen-rich clumps, the \emph{R1, R2, B1, B2, B3} and \emph{B4} structures can be readily identified. The so-called \emph{run-away} or \emph{RK} knot can be easily spotted to the SE (lower left), close to the edge of field 4.

\subsection{Comparison with X-ray observations}\label{Sec:xray}

\begin{figure*}[htb!]
\centerline{\includegraphics[scale=0.6]{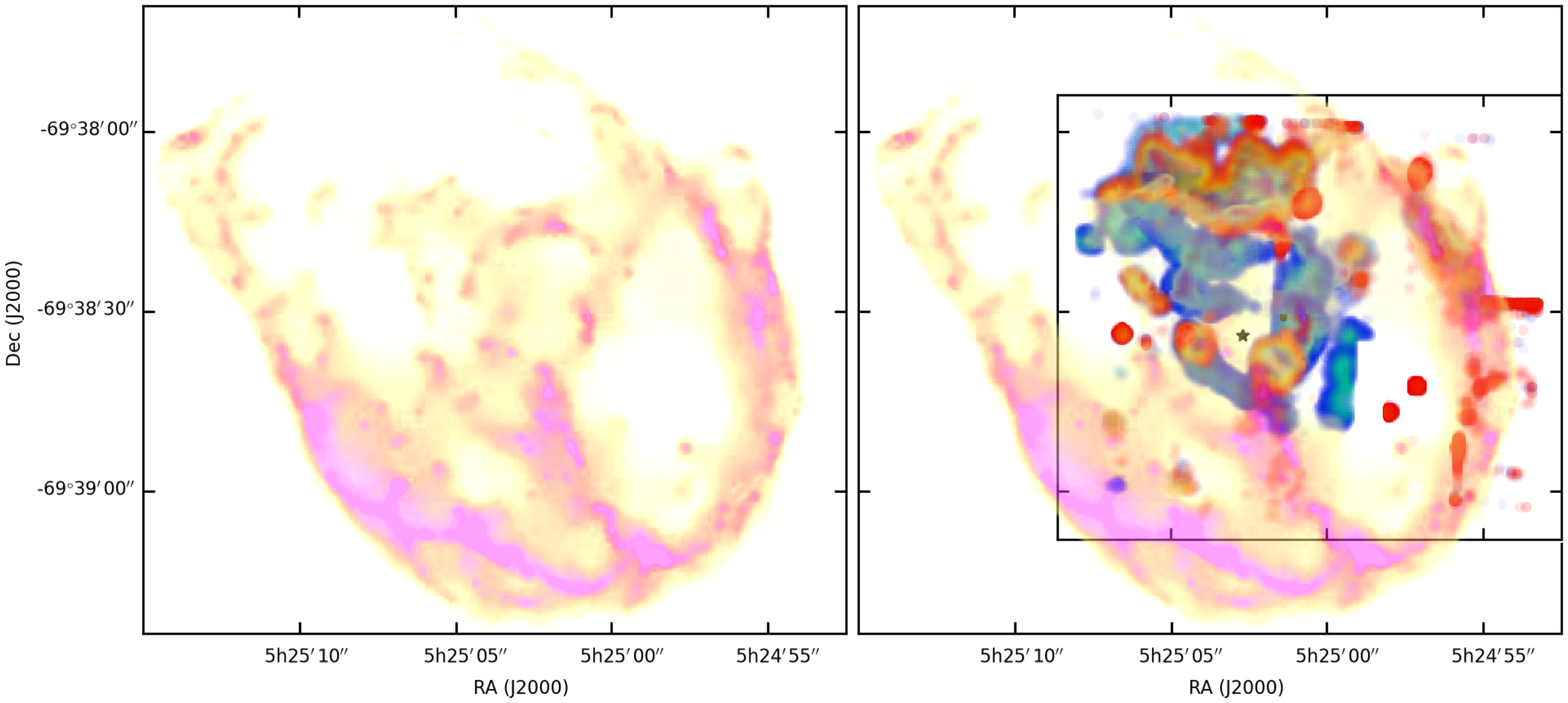}}
\caption{Left : Chandra X-ray observation of N132D, from \cite{Xiao08}. Right : idem, but with the X-ray overlaid over the [O~III] (blue) and H$\beta$ (red) emission from Fig.~\ref{fig:above}-right.}\label{fig:xray}
\end{figure*}

The X-ray data and the comparison of the X-ray image with the positions of the oxygen-rich filaments provides extensive information on the location of the forward and reverse shock and on the overall structure of the SNR. The outer rim of the X-ray image defines the outer blast wave, and local enhancements in the X-ray emissivity associated with oxygen-rich knots trace the position of the reverse shock.  In Fig.~\ref{fig:xray}, we compare the optical [O~III] $\lambda$ 5007 and H$\beta$ emission to a recent \emph{Chandra} X-ray observation of SNR N132D. On the left plot, we show the X-ray data from \cite{Xiao08} (Fig.~2 in their paper). Strong X-ray emission is displayed in purple while weaker emission is in yellow. The characteristic shape of the outer ellipse of emission, reminiscent of a horseshoe, oriented in a NE-SW direction is easily seen. On the right hand panel, Fig.~\ref{fig:above}-Right,  we overlay the X-ray emission over the [O~III] $\lambda$ 5007 and H$\beta$ emission. 

Several observations can be made. First, increased X-ray emissivity along the outer NW edge of the remnant is associated with enhanced optical emission. This feature is associated to the position of the forward (blast-wave) shock and shocked ISM \citep[][]{Morse96}. The optical enhancement shows that the ISM shocks are already becoming radiative. A similar correlation between X-ray and optical emission  is also seen in the southern rim, but this region is not completely covered by our observations. However, we do see two faint optical knots in the SE portion of the outer blast-wave region, as well as the Runaway Knot (RK), identified previously in Fig.~\ref{fig:above}. The abundance of oxygen and lack of H$\beta$ emission, as well as its high radial velocity ($\sim$ 1000 km~s$^{-1}$) does suggest that it is a \emph{bona fide} O-rich clump. This clump is also associated with a large local enhancement in the X-ray emission. However, its position, away from any other oxygen-rich clumps, and its possible interaction with the forward blast wave is curious. The nature of this knot will be discussed further in Sec. \ref{Sec:RK}.

Second, the inner X-ray ring is almost perfectly superposed to the O-rich knots. \cite{Borkowski07} who observed an enhanced emission of oxygen in the X-ray spectrum at those location had already suggested it to be possibly the location of the reverse shock wave, or potentially be the consequences of a Ni-rich bubble. Clearly, its close association with O-rich material makes the reverse shock hypothesis the most likely one.

Third, the \emph{B1} knot, to the SW of the SNR center, appears not to be associated with any X-ray emission, unlike the other O-rich clumps. This is somewhat of a mystery, since we would expect  some X-ray emission to be associated with it. It is remotely possible that these knots are being photoionized by the diffuse X-ray field. The intensity of its emission is slightly smaller than the other O-rich clumps (see Fig.~\ref{fig:3dproj}).

Fourth, the \emph{Lasker's Bowl} feature fits almost perfectly into a hole in the X-ray emission, This suggests that this dense cloud in the ISM lies in the foreground, and has sufficient column density to absorb the underlying diffuse X-ray emission. It is delineated on part of its southern boundary by an X-ray enhancement, presumably generated by the stand-off shock associated with the cloud.

\section{The 3D map of the ejecta}\label{Sec:3d}
In this Section, we use the idea developed by \cite{Vogt10a} to reconstruct a 3D map of the oxygen rich ejecta in SNR N132D. Assuming that N132D is still in its free expansion phase implies that there exist a linear relation in between the radial velocity $v_r$ and the distance $z$ covered by the clumps since the explosion. The scale depends on two parameters : the distance $D$ to SNR N132D, and the age of the explosion. N132D is located in the LMC which makes the distance determination easier. Here, we will assume $D$=50~kpc \citep[][]{Panagia91,Marel02}. Determining the age of the explosion, however, proves less straight forward than the case of the earlier studied 1E 0102.2-7219 in the SMC. Currently, there are no proper motion measurements of the expansion rate of the ejecta. The only age estimates are based on size and radial velocity measurements. Since these require some kind of assumption about the 3D shape of the SNR, they are much more uncertain. Current age estimates range from $\sim$1300 years \citep[][]{Lasker80} to $\sim$3000 years and above \citep[][]{Morse95}. Let us first explore the consequences of making different age estimates on the resulting 3D map, before studying in detail our best estimate of the  3D shape of N132D in Section~\ref{Sec:ring}.

\subsection{Age estimates of N132D}

\begin{figure*}[htb!]
\centerline{\includegraphics[scale=0.6]{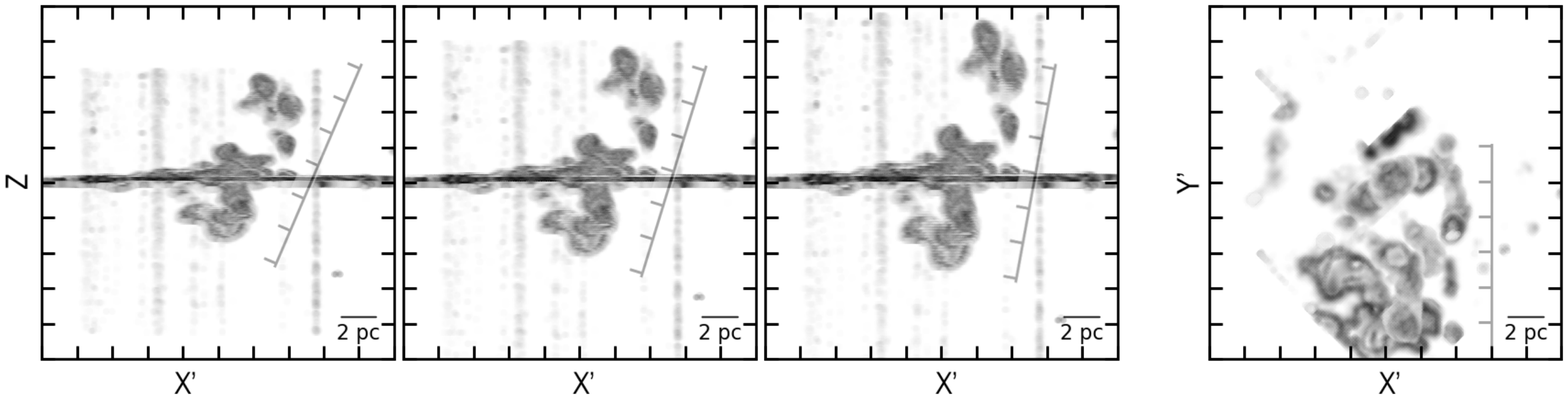}}
\caption{Left : 3D map of the O-rich ejecta in N132D, projected along the Y' axis, where the Z axis is the line of sight, and Y' the direction perpendicular to Z and in the plane of the ring-shaped ejecta, assuming an age of 2000, 2500 and 3000 years (left to right). Right : top view of the ejecta in a X'Y' coordinate system.  X'Y'Z forms an orthonormal base, and the Earth is to the top of the pictures. The linear scale is given in each panel, assuming a distance to the LMC of 50 kpc \citep[e.g.][]{Marel02}. Each tick corresponds to 2~pc. For the projection on the X'-Y' plane, the low velocity structures, with  -100 km s$^{-1}$ $\leq$ v$_r$ $\leq$ 100 km s$^{-1}$ have been skipped to avoid filling the figure. See in the text for details.}\label{fig:angle}
\end{figure*}

As we discuss in the next Section, the bulk of the oxygen-rich ejecta trace out a  ring tilted with respect to the line of sight, a model that is broadly in agreement with that of  \citet{Lasker80}. Taking the distance to the LMC to be $D=50$~kpc, the effect of different age estimates will be to either expand or contract the ejecta along the line of sight (z-axis). In  Fig.~\ref{fig:angle}, we have projected the 3D map on an X'Y'Z coordinate system, where Z is the line of sight direction, with the Earth to the top of the Figure, and Y' is in the ejecta ring plane. In the three left hand panels of this figure, we plot the projection of the 3D map assuming an age of 2000, 2500 and 3000 years (from left to right), and on the right hand panel the view in the plane of the sky, which obviously does not depend on the age estimate. 

Assuming that the central ring is indeed a true ring, its projected size in the X'-Z plane should equal its long axis in the X'-Y' plane (or equivalently, the  X-Y plane). This can be estimated in Fig.~\ref{fig:angle}. In the top view, we have plotted a 12~pc ruler in grey shades, which we find to be the length of the long axis of the ejecta in this plane. We did not take the \emph{B1} knot in consideration in this process, since it may lie beyond the reverse shock. We have then plotted the same linear scale of 12~pc in a grey shade on every three projections on the X'-Z plane (left panels ). Clearly, the 2000 years estimate has a ring too small by $\sim$2~pc, while on the other hand the 3000 years estimate ring is too big by $\sim$2~pc. The central estimate of 2500 years has a ring that extends over the same distance, $\sim$12~pc,  in the X'-Z plane as it does in the X'-Y' plane.  With this age estimate, the ring will then be a true circle, with a radius of $\sim$12~pc, and tilted by $\sim 25$ degrees with respect to the line of sight. Thus, an age of  2500 years is our preferred estimate, which will be used in the remaining Sections of this paper. We should also mention here that  \cite{Lasker80} found a good fit to the ejecta shape assuming a ring tilted at $\sim 45$ degrees, which is consistent with our results, bearing in mind that his age estimate for the SNR was  only $\sim 1300$ years.

\subsection{The ring-like structure}\label{Sec:ring}

\begin{figure*}[htb!]
\centerline{\includegraphics[scale=0.7]{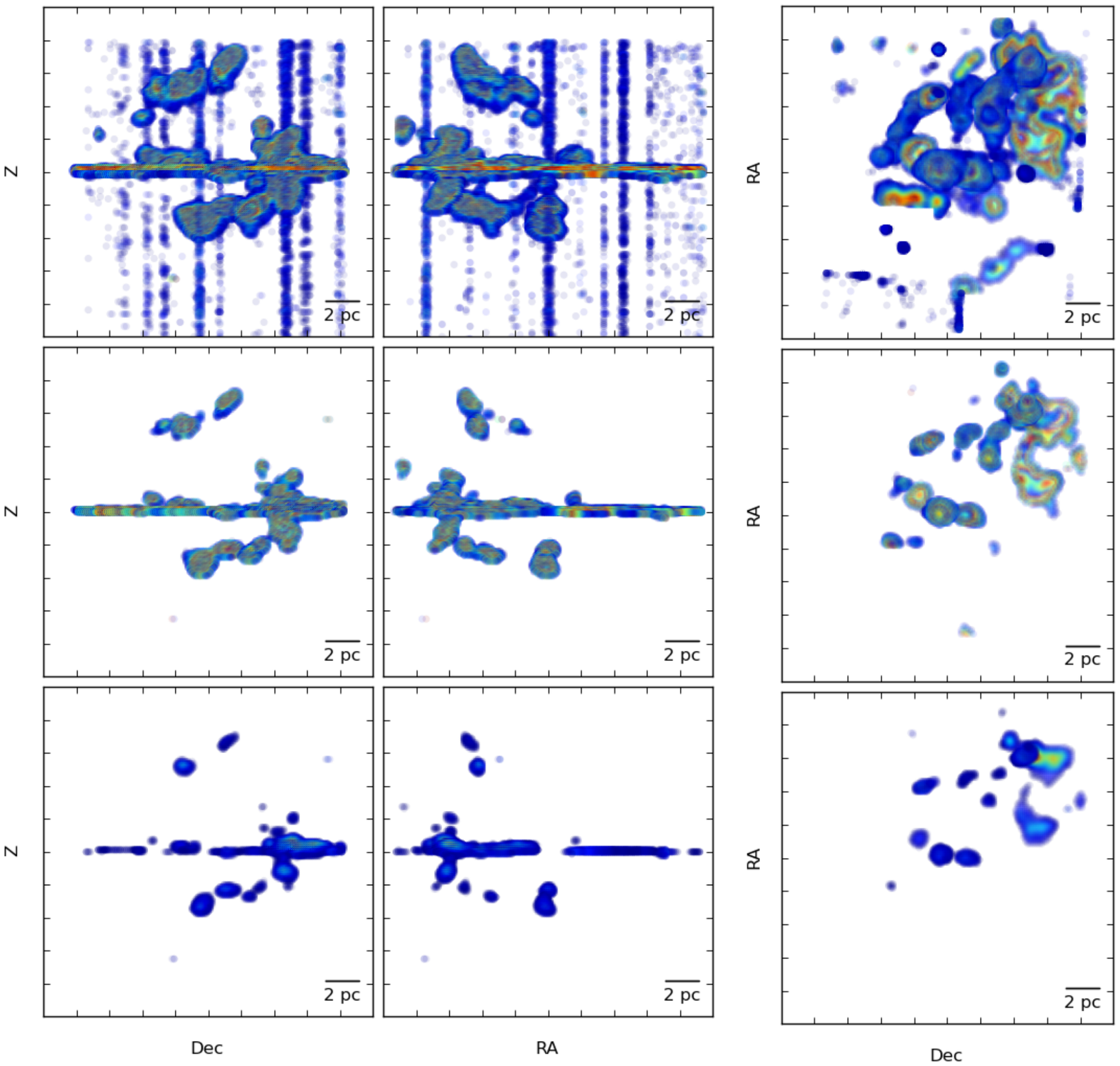}}
\caption{Projections of the data cube along the line of sight (Z), the RA and Dec directions. The rainbow color ramp is linear in the number of counts per pixels, and ranges from 18-60, 60-120 and $\geq$120 for the top, middle and bottom row respectively. For the projection along the Z axis, the low velocity structures, with  -100 km s$^{-1}$ $\leq$ v$_r$ $\leq$ 100 km s$^{-1}$ have been skipped to avoid filling the figure. Earth is to the top of the pictures. The linear scale is given in each panel, assuming a distance to the LMC of 50 kpc \citep[e.g.][]{Marel02}, and an age of 2500 years for N132D. Each tick corresponds to 2~pc, and at the estimated age and distance, $v_r$=1000 km~s$^{-1}$ corresponds to 2.55~pc, or 1~pc corresponds to 391.39 km~s$^{-1}$ in radial velocity. }\label{fig:3dproj}
\end{figure*}

Adopting an age of $T$=2500 years for N132D, and a distance $D$=50~kpc \citep{Panagia91}, gives the following transformation to our data cube axis units: \begin{eqnarray}
v_r \textrm{ [km s}^{-1}\textrm{]}\rightarrow z \textrm{ [pc]}=v_r\cdot T \nonumber \\
x \textrm{ [arc sec]}\rightarrow x\textrm{ [pc]}=x\textrm{ [arc sec]}\cdot D \nonumber\\
y \textrm{ [arc sec]}\rightarrow y\textrm{ [pc]}=y\textrm{ [arc sec]}\cdot D 
\end{eqnarray}The resulting data cube provides the 3D map of the oxygen-rich ejecta in SNR N132D. In Fig.~\ref{fig:3dproj}, we present side, front and top projections of this 3D map. The rainbow color ramp is linear in the number of counts per pixels, and ranges from 18-60, 60-120 and $\geq$120 for the top, middle and bottom rows respectively. For the projection along the Z axis, the low velocity data, with  -100 km s$^{-1}$ $\leq$ v$_r$ $\leq$ 100 km s$^{-1}$ have been skipped to avoid filling the figure with emission which would obscure the salient structural features. For the side and front projections, the Earth is to the top of the plot. A 2~pc scale is shown for information, which is also the inter-ticks distance. At the estimated age and distance of N132D, $v_r$=1000 km~s$^{-1}$ corresponds to 2.55~pc, or 1~pc corresponds to 391.39 km~s$^{-1}$ in radial velocity. In the front and side views, the vertical stripes are due to field stars, while the horizontal plane is due to the low velocity gas in SNR N132D.

The ejecta have an approximately circular symmetry in the side and front views. In comparison, in the top view, they seem to form an ellipse with major axis in the NE-SW direction. This suggests that the ejecta have the shape of a tilted ring. Note that the blue-shifted material has a higher velocities than the red-shifted material. This results in a mean radial velocity of the oxygen-rich knots which is blue-shifted with respect to the systemic velocity of the surrounding ISM by several hundreds of kilometers per second. This has been noticed in previous studies \citep[][]{Lasker80,Sutherland95a}. 

The ISM knots, previously identified in Section~\ref{sec:orich}, display smaller radial velocities, of the order of a few hundred km s$^{-1}$. These are disposed in a ring pattern in the plane of the sky. The resulting shape of the O-rich emission, using our 3D reconstructing technique, is that of a double interleaved ring system, one being composed of O-rich material, and the other being composed of shocked ISM knots.The fact that many ISM knots happen to coincide with O-rich knots when viewed from above is probably a coincidence, since these are shocked by the blast wave, not the oxygen-rich material passing through the reverse shock.  

However, there may be a case to be made for a closer interaction between either the \emph{R1} or the  \emph{B4} clumps and the Lasker's Bowl complex, since their combined shape matches perfectly to the overall shape of the Lasker's Bowl feature, suggesting a direct interaction. However, there are two issues arising from this scenario. First, the \emph{B4} and \emph{R1} clumps cannot be related to each other, one being blue-shifted, and the other red-shifted, and therefore well-separated in space. Second, the Lasker's Bowl appears to be blue shifted. This is consistent with it being a foreground object, as suggested by the X-ray shadowing, since the cloud shock within the Lasker's Bowl complex would be propagating through it towards the observer. The \emph{R1} clump cannot be hitting it as is moving in the other direction. However, there is still a case to be made that the \emph{B4} clump is associated, as it lies in a region of enhanced X-ray emission, possibly associated with the reverse (bow) shock generated when the blast wave over-ran the rear surface of the large ISM cloud.

\begin{figure*}[htb!]
\centerline{\includegraphics[scale=0.6]{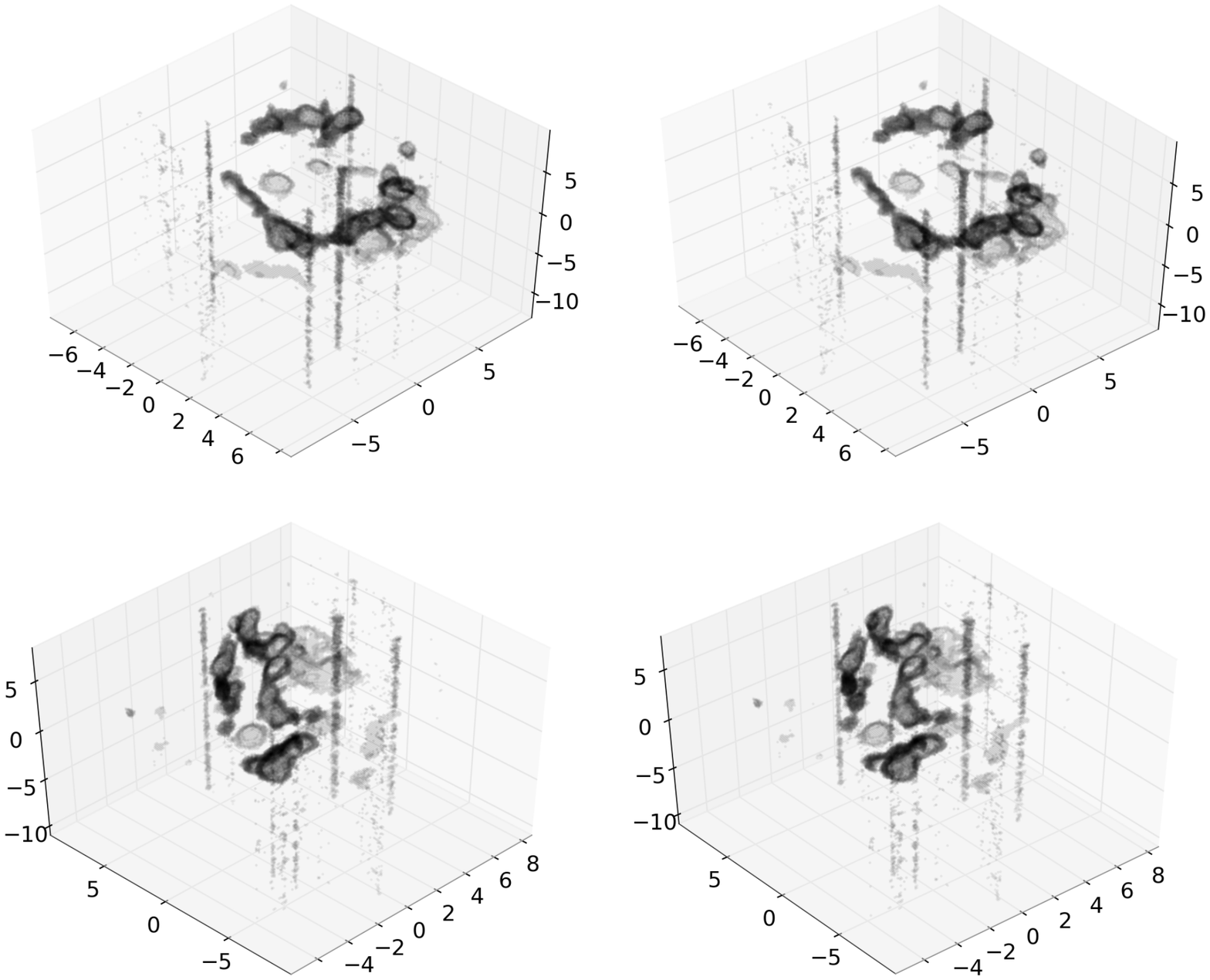}}
\caption{Top : 3D stereoscopic maps of the oxygen rich ejecta in N132D, viewed from the NW and +45$^{\circ}$ elevation with respect to the plane of the sky.  Bottom : the same, but this time viewed from the SW. Both pairs show the weakly emitting filaments in the accumulated signal range of $18-60$ counts per pixel. The scales are given in pc, assuming a distance to the LMC of 50~kpc and an age of 2500 years for N132D. }\label{fig:3D}
\end{figure*}

As a complement to the projection of our 3D map in Fig.~\ref{fig:3dproj}, we have created stereo pairs\footnote{ \emph{Stereo pairs} or \emph{stereograms} enable us to see a Real 3D image of an object using two 2-D images. The two projections are separated by 5 degrees in azimuth. The top pair looks at N132D from the NW and an elevation of +45 degrees, while the bottom pair is from a SW and +45 degrees point of view.
By crossing ones eyes, it is possible to reconstruct  3D images of the object by combining the two projections.}.  A similar technique has been used for SNR 1E 0102.2-7219 by \cite{Vogt10a}, and a detailed description of the method, and its advantages, can be found in \cite{Vogt10b}. Two stereo pairs of the oxygen rich ejecta in SNR N132D are shown in Figure~\ref{fig:3D}. In those images, the ring-shaped structure of the oxygen-rich material becomes very apparent. The ISM knots appear as second loop in the plane of the sky. However, this is simply because the cloud shocks which excite the ISM clouds are of low velocity. This structure is therefore an artifact, having nothing to do with the real 3D structure of the ISM clouds. 

In order to give a better feeling for the ring shape of the O-rich ejecta, which highlights its distorted aspect on the red-shifted side, we have also created two interactive maps and a stereo movie that can all be freely downloaded here : \newline \textsf{http://hdl.handle.net/102.100.100/4451}. \newline The interactive maps are based on the concept described by \cite{Barnes08}, and are similar to that of SNR 1E 0102.2-7219 in \cite{Vogt10a}. We have used \texttt{Python} and \texttt{MayaVi} in order to generate a CAD-like\footnote{CAD=Computer-Aided Design} 3D map of the ejecta, and \texttt{Adobe Acrobat 9 Pro extended} to produce a .pdf of it. In order to be viewed, the only software needed is an up-to-date version (v.8 or above) of \texttt{Adobe Acrobat Reader}\footnote{\textsf{http://get.adobe.com/reader/ }}. The user is able to interact in several ways with the resulting file : rotate around the ejecta, or zoom in and out at will, for example. A handy cross-section tool is also available which enables the data to be sliced into planes. 

The data and the color code used in those 3D maps are identical to the top row of Fig.~\ref{fig:3dproj}. It represents the low intensity ejecta, with the number of counts per pixel ranging from 18 - 60. Note that the low velocity gas, with -100 km s$^{-1}$ $\leq$ $v_r$ $\leq$ 100 km s$^{-1}$ is not plotted in order to avoid filling the maps, and obscuring the high-velocity details. Each pixel is represented by a circle (and not a sphere) of fixed radius in order to keep the size of the file as small as possible. As a result, the ejecta are slightly transparent in this representation. Three red-green-blue axes, located on the bottom left of the map, enable to orientate the map in the 3D space. The red-X axis corresponds to the Dec direction, while the green-Y axis corresponds to the RA direction, and the blue-Z axis corresponds to the line of sight, with the Earth in the positive direction. In the map called \textsf{N132D\_3d\_OIII\_Hb.pdf}, the H$\beta$ emission as been plotted in red over the blue [O~III] emission (shown alone in the map \textsf{N132D\_3d\_OIII.pdf}), similarly to the data shown in Fig.~\ref{fig:above}-Right. As for the Real 3D movie, it can be looked at using the stereo pairs viewing technique described above and in \cite{Vogt10b}. Note that the plotted data is identical to that contained in the interactive 3D map.

\section{Discussion}\label{Sec:RK}

In its broad characteristics, N132D is a typical young SNR, with several high-velocity oxygen-rich filaments, excited by a reverse blast wave, while the forward shock wave is expanding into space, shocking the surrounding medium. The forward shock wave is nicely traced by strong X-ray emission of the shocked ISM, a feature which can be easily observed in the case of N132D. The reverse shock wave signature, in that of X-ray emission superposed to the oxygen-rich ejecta,  has only been recently detected \citep[][]{Borkowski07,Xiao08}.  The Ni-bubble scenario for this feature \citep[][]{Socrates05,Wang05}, as suggested by \cite{Borkowski07}, appears rather unlikely. This scenario might be responsible for initially shaping the O-rich ejecta into a ring, but not for the emission of X-rays 2500 years after the shock breakout.

The shape of the ejecta within N132D has been a puzzling question in the past studies of this remnant. From our 3D maps, it is now clear that the ejecta traces a distorted ring, as first suggested in the pioneering work of \citet{Lasker80}. However, with the greater age of $\sim 2500$ years that we have inferred here, the ring is less inclined that Lasker initially thought, lying at $\sim$20-30 degrees with respect to the line of sight. The ring appears rather distorted on the redshifted side, and is systematically blue-shifted by $\sim$400 km s$^{-1}$ with respect to the LMC rest frame. This fact, already identified in previous studies, is most probably associated with an asymmetry during the explosion. 

The ratio of the forward shock radius to the reverse shock radius is $\sim 3.20$. This high value is a sign that N132D is in rather a late stage for  a young SNR. The reverse shock must have already started accelerating towards the center, at which point the SNR will complete its transition to the Sedov phase  \citep[][]{reynolds2008}. In this scenario, the shocked ejecta, moving at ballistic velocities since the explosion should be expanding at comparatively slower speed with respect to other younger SNR. 

Given the fact that the oxygen-rich ejecta form a (somewhat thick) ring, and since these ejecta are amongst the slowest of the helium burnt products of the exploding star, it is very tempting to propose that these represent a denser equatorial ring of ejecta from an initially rotating star. In this scenario the polar ejecta will have higher velocities, but lower column densities, and so will have interacted with more of the surrounding ISM, allowing the reverse shock to fully propagate through them.

It is interesting that the major axis of the outer X-ray shell associated with the blast wave also forms an ellipse with $PA\sim 45$~deg., and is more or less aligned with the major axis of the fast-moving ring of ejecta in the XY plane (in Fig.~\ref{fig:angle}, the X'Y'Z reference frame is rotated by $45$~deg. around Z with respect to the XYZ reference frame.). The outer shell may be either itself a ring-like structure, the dense mid-plane of an oblate ellipsoid or (less probably) it may represent a ßattened (oblate) ellipsoid (or pumpkin shape). 

If interpreted as a ring, and assuming that, as for the oxygen-rich ejecta, this ring is circular, then we infer that it has an inclination to the line of sight of $\sim 40$~deg., as deduced from the length ratio of the primary to the secondary axis of the outer ellipsoid X-ray shell.  This is sufficiently close in angle to the $\sim 25$~deg. inferred for the oxygen-rich ring to speculate that the outer X-ray ring may represent the remnants of a (possibly toroidal) ring nebula ejected during an earlier Wolf-Rayet phase of evolution of the SN progenitor. This hypothesis is capable of a number of observational tests. First, if the optical outer shell represents material originally ejected from the central star, then we would expect it to be enhanced in the products of hydrogen-burning, notably helium and nitrogen. \cite{Blair00} have obtained the optical spectra of one of the bright knot within the outer ring, and found its elements abundances to be similar to that of the ISM in the LMC. This analysis would need to be expanded to other locations on the outer rim by further spectrophotometric work in order to detect any helium or nitrogen enhancement. Second, we would expect the blast wave to have interacted with this ring already, and to have swept around it. Indeed, such appears to be the case, since the radiative small clumps of gas in both the NW and  and SE of the remnant clearly lie inside outer X-ray bright shell (see for example Figure \ref{fig:xray}). Third, in this scenario, the polar direction will have mostly been swept clear of interstellar gas by the fast wind. This will have the effect that, contrary to our supposition above that the reverse shock has propagated back to the center of the remnant in the polar direction, the development of the reverse shock is relatively less advanced in the polar direction due to the smaller column of ISM intercepted in this direction. 

This idea can be directly tested, as in this case, we should hope to detect very fast oxygen-rich ejecta moving out in the NW and SE sides of the SNR.
The run-away knot may well provide an example of this phenomenon. This knot has only a modest radial velocity $v_r\cong850$ km~s$^{-1}$, but is located surprisingly far away from the SNR center, and also a long way from the other oxygen-rich clumps. However, its strong [O~III] emission and lack of any hydrogen recombination line emission classifies it at a \emph{bona fide}  oxygen-rich knot. Its location ($\sim$8~pc from the center) implies a transverse velocity of $\sim3130$ km~s$^{-1}$ and a total space velocity of $\sim$3250 km~s$^{-1}$. This value is in striking contrast with the other oxygen-rich knots. In projection, the Run-Away knot is coincident with a local enhancement in the X-rays, and lies close to the forward shock as traced by the X-ray ring or shell. The position of this Run-Away knot, almost perpendicular to the axis of the distorted ring, and its high space velocity make the idea that it is part of a fast jet-like polar ejection quite appealing. Such evidence of jets, highlighted by the shape of the ejecta, has been seen in Cas A \citep[][]{Fesen96,Fesen06a}. Many simulations have also attempted to create such jets  \citep{Khokhlov99,MacFadyen01}. The lack of any other \emph{polar-jet} clumps is possibly because of their small size and/or faintness. The possibility of a polar jet is also supported by the fact that there is a NW-SE bar-like enhancement in both the the X-ray emission and in the radio emission  \citep[][]{Dickel95} which is well aligned with the position angle of the run-away knot relative to the centre of the SNR. This is clearly seen in Figure \ref{fig:xray}. It would be useful to obtain second-epoch HST images of this remnant to discover the transverse velocity of the run-away knot, and to search for other very faint and small oxygen-rich clumps.

\section{Conclusions}\label{Sec:concl}
In this paper we have applied the same analysis techniques to the integral field spectrometry of N132D as we earlier did to the ejecta in SNR 1E 0102.2-7219 \citep[][]{Vogt10a}. However the results are strikingly different. Although clearly more evolved than SNR 1E 0102.2-7219, N132D has evidently had a quite different evolutionary history. While SNR 1E 0102.2-7219 showed evidence for bipolar ejection, N132D on the contrary shows strong evidence for equatorial ejection either before the supernova event and/or at the time of explosion. This produced the overall X-ray morphology, and we currently see an expanding ring of oxygen-rich material currently passing through the reverse shock, a model very similar to that originally advocated by \citet{Lasker80}. This ring is broadly aligned with a larger outer ring or toroid of pre-supernova ejecta traced by the bright X-ray emitting region. We have also found evidence for a polar jet of material visible in the X-ray and radio emission, and associated with a single high-velocity oxygen-rich knot. The existence of a high-velocity polar jet is reminiscent of what is seen in Cassiopeia A \citep{Fesen06a,Fesen06b}, although in the case of N132D the angle of inclination of the jet to the line of sight is smaller. The formation of such jets requires particular conditions \citep{vanVeelen09,Schure08} may well be related to the phenomenon of the Gamma-Ray Burst sources (GRBs).

\acknowledgments

Dopita acknowledges the support of the Australian Research Council (ARC) through Discovery  projects DP0984657 and DP0664434.  This research has made use of data products from the Two Micron All Sky Survey, which is a joint project of the University of Massachusetts and the Infrared Processing and Analysis Center/California Institute of Technology, funded by the National Aeronautics and Space Administration and the National Science Foundation. It has also used the NASA/IPAC Extragalactic Database (NED) which is operated by  the Jet Propulsion Laboratory, California Institute of Technology, under contract with the National  Aeronautics and Space Administration.  This research has also made use of NASA's Astrophysics Data System, and SAOImage DS9, developed by the Smithsonian Astrophysical Observatory.

\end{document}